\documentclass[conference]{IEEEtran}
\usepackage{amsmath,amsfonts,amssymb,mathtools,dsfont,multicol,multirow,hhline,diagbox}
\usepackage{graphicx}
\usepackage{algorithmic,algorithm,threeparttable}
\usepackage{amsthm}
\usepackage[noadjust]{cite}

\newtheorem{theorem}{Theorem}
\newtheorem{lemma}{Lemma}
\newtheorem{prop}{Proposition}
\newtheorem{example}{Example}

\ifCLASSINFOpdf
\else
\fi
\IEEEoverridecommandlockouts
\begin{document}
\title{A Tractable Framework for Performance Analysis of Dense Multi-Antenna Networks}
\author{\IEEEauthorblockN{Xianghao Yu$^*$, Chang Li$^{\dag}$, Jun Zhang$^*$, and Khaled B. Letaief$^{*\ddag}$, \emph{Fellow, IEEE} }
\IEEEauthorblockA{$^*$Dept. of ECE, The Hong Kong University of Science and Technology, Hong Kong\\
	$^\dag$National Institute of Standards and Technology, Gaithersburg, MD\\
	$^\ddag$Hamad Bin Khalifa University, Doha, Qatar\\
Email: $^*$\{xyuam, eejzhang, eekhaled\}@ust.hk, $^\dag$chang.li@nist.gov, $^\ddag$kletaief@hbku.edu.qa}
\thanks{This work was supported by the Hong Kong Research Grants Council under Grant No. 16210216. 
}
}

\maketitle

\begin{abstract}
Densifying the network and deploying more antennas at each access point are two principal ways to boost the capacity of wireless networks.
However, due to the complicated distributions of random signal and interference channel gains, largely induced by various space-time processing techniques, it is highly challenging to quantitatively characterize the performance of dense multi-antenna networks.
In this paper, using tools from stochastic geometry, a tractable framework is proposed for the analytical evaluation of such networks. 
The major result is an innovative representation of the coverage probability, as an induced $\ell_1$-norm of a Toeplitz matrix. This compact representation incorporates lots of existing analytical results on single- and multi-antenna networks as special cases, and its evaluation is almost as simple as the single-antenna case with Rayleigh fading. 
To illustrate its effectiveness, we apply the proposed framework to investigate two kinds of prevalent dense wireless networks, i.e., physical layer security aware networks and millimeter-wave networks.
In both examples, in addition to tractable analytical results of relevant performance metrics, insightful design guidelines are also analytically obtained.
\end{abstract}


\IEEEpeerreviewmaketitle

\section{Introduction}
To meet the ever-increasing mobile data traffic explosion, there is a tremendous demand in boosting the capacity of wireless networks. One promising way is to exploit the spatial domain resources by deploying more antennas at transceivers, especially at the base station (BS) side, e.g., via the recently emerged ``Massive MIMO" technique \cite{6375940}.
Another effective way to increase the network capacity is via network densification \cite{7010535}, which can significantly improve the area spectral efficiency (ASE). However, to design and evaluate dense multi-antenna networks is a highly challenging task, which may hinder their wide deployment.

The main difficulty to analytically characterize the network-level performance comes from the complicated signal and interference distributions, which depend on the applied multi-antenna transmission strategy, as well as the channel model. Previous studies have revealed that the gamma distribution is typically encountered when evaluating various multi-antenna systems. 
For example, it was shown in \cite{4712724,6775036} that with Rayleigh fading the channel gain for the information signal is gamma distributed under different multi-antenna transmission techniques, e.g., zero forcing (ZF) and maximal ratio transmission (MRT) beamforming. 
For more general multi-antenna transmission strategies, gamma distribution was shown to be an accurate approximation of the channel gain \cite{5953530}. Furthermore, Nakagami fading will generally lead to a gamma distributed channel gain.
 While existing results are mainly for the Rayleigh fading scenario, i.e., with exponentially distributed channel gains, an analytical framework that can effectively handle gamma distributed channel gains is highly desirable for studying dense multi-antenna networks. On the other hand, with network densification, the distribution of the aggregated interference becomes intricate, which brings additional challenges to the performance evaluation. A random network model based on Poisson point processes (PPPs) has been adopted extensively to model the dense BS deployments. With the help of stochastic geometry, this model turns out to be tractable and can effectively characterize the aggregated interference \cite{6042301}.

There have been some attempts to analytically evaluate multi-antenna wireless networks based on the random network model \cite{4712724,5673756,5351444,6932503}. 
Taylor expansion was used in \cite{4712724} for approximating the interference power distribution in ad hoc networks. 
Analytical expressions provided in \cite{5673756,5351444} were in complicated forms via many special functions, e.g., Bell polynomials and beta functions.
A more recent work \cite{6932503} adopted an upper bound for the cumulative probability function (cdf) to handle the gamma distributed channel gains, which led to a closed-form expression for the coverage probability.
Unfortunately, the available results, typically with approximations, are all in complicated forms, which cannot yield further insights for network design and optimization.

Recently, some promising results were produced in our previous works \cite{6775036,7038201,7412737}, where closed-form expressions were derived for various performance metrics in multi-antenna heterogeneous networks.
These results disclosed the potential of yielding a systematic way to analyze multi-antenna networks, and provided design guidelines for some specific network models and multi-antenna transmission techniques.
In this paper, we shall extend the analyses in \cite{6042301,6775036,7038201,7412737} to a more general framework, which is applicable to networks where the signal channel gain is assumed to be gamma distributed while the interference channel gains are with arbitrary distributions.
In particular, the recursive relations between the $n$-th derivatives of the Laplace transform are exploited, based on which a novel representation of the coverage probability is derived, i.e., an induced $\ell_1$-norm of a Toeplitz matrix representation.
With the proposed framework, the complexity of evaluating dense multi-antenna networks becomes comparable to the single-antenna case. Moreover, many analytical techniques developed for conventional single-antenna networks can be easily transplanted to the general multi-antenna setting.

To illustrate its effectiveness, the proposed framework is then applied to two example networks, i.e., physical layer security aware networks and millimeter-wave (mmWave) networks, for which fewer analytical results are available.
With the new analytical tool, we are able to derive a new set of tractable results for these networks. With these results, we also investigate two critical design problems, i.e., the trade-off between the jamming and interference nulling in security aware networks, as well as the impact of the array size in mmWave networks.

\section{A Unified Analytical Framework}\label{II}
\subsection{Analytical Framework for Multi-Antenna Networks}
Consider a dense multi-antenna wireless network, where the spatial locations of transmitters are modeled as a homogeneous PPP, denoted as $\Phi$ in $\mathbb{R}^2$ with density $\lambda_\mathrm{t}$. Each transmitter communicates with multiple single-antenna receivers with fixed transmit power. 
We focus on the performance analysis of the typical receiver at the origin, and the signal-to-interference-plus-noise ratio (SINR) is given by
\begin{equation}\label{SINR1}
\mathrm{SINR}=\dfrac{g_{x_0} r_0^{-\alpha}}{\sigma_\mathrm{n}^2 + \sum_{x\in\Phi^\prime}g_x \|x\|^{-\alpha}},
\end{equation}
where $r_0=\Vert x_0\Vert$ is the distance from the typical receiver to its associated transmitter located at $x_0$, with the probability density function (pdf) $f_{r_0}(r)$. The noise power is normalized, depending on the system setting, and is denoted as $\sigma_\mathrm{n}^2$. The channel gains for the information signal and interference from the transmitter located at $x$ are denoted as $g_{x_0}$ and $g_{x}$, respectively. The signal channel gain $g_{x_0}$ is gamma distributed, i.e., $g_{x_0}\sim\mathrm{Gamma}(M,\theta)$, where $M$ and $\theta$ are shape and scale parameters of the gamma distribution. We assume $(g_x)_{x\in\Phi^\prime}$ is a family of independent and non-negative random variables with arbitrary distributions.
The locations of the concerned interfering transmitters are denoted as $\Phi^\prime$, which can be composed of any PPP conditional on $x_0$. In particular, $\Phi^\prime$ can be a union of several different types of interferers that are distributed according to different PPPs  $\Phi_j^\prime$, and each type of interferer has different densities $\lambda_{\mathrm{t},j}$ and interference channel gains $g_{x,j}$.

We focus on the coverage probability, defined as
\begin{equation}\label{coveragedef}
p_\mathrm{c}(\gamma)=\mathbb{P}(\mathrm{SINR}>\gamma),
\end{equation}
where $\gamma$ denotes the SINR threshold. Many other typical network performance metrics, e.g., ASE, average throughput, and energy efficiency, can be analyzed based on the results for the coverage probability \cite{6775036,7038201,7412737,haenggi2012stochastic}.

In this section, we will provide a unified analytical framework for dense multi-antenna wireless networks.
First, the coverage probability defined in \eqref{coveragedef} can be written as
\begin{equation}
p_\mathrm{c}(\gamma)
=\mathbb{P}\left[g_{x_0}>\gamma r_0^\alpha\left(\sigma_\mathrm{n}^2 +I\right)\right],\label{coverageprob}
\end{equation}
where $I\triangleq\sum_{x\in\Phi^\prime}g_x \|x\|^{-\alpha}$.
As mentioned before, one main difficulty of the analysis comes from the gamma distributed random variable $g_{x_0}$. Different from previous works that adopted approximations \cite{4712724,6932503}, in this paper, we will derive a compact and exact expression for this probability. According to the cdf of gamma distribution, the coverage probability \eqref{coverageprob} is firstly rewritten as
\begin{align}
p_\mathrm{c}(\gamma)&=\mathbb{E}_{r_0}\left\{\sum_{n=0}^{M-1}\frac{(\gamma r_0^\alpha/\theta)^n}{n!}\mathbb{E}_I\left[(\sigma_\mathrm{n}^2+I)^ne^{-\frac{\gamma r_0^\alpha}{\theta} (\sigma_\mathrm{n}^2+I)}\right]\right\}\nonumber\\
&=\mathbb{E}_{r_0}\left[\sum_{n=0}^{M-1}\frac{(-s)^n}{n!}\mathcal{L}^{(n)}(s)\right],\label{eq4}
\end{align}
where $s\triangleq\gamma r_0^\alpha/\theta$, $\mathcal{L}(s)=e^{-s\sigma_\mathrm{n}^2}\mathbb{E}_I\left[e^{-sI}\right]$ is the Laplace transform of noise and interference. The notation $\mathcal{L}^{(n)}(s)$ stands for the $n$-th derivative of $\mathcal{L}(s)$.
According to the probability generating functional (PGFL) of PPP,
the Laplace transform $\mathcal{L}(s)$ can be expressed in a general exponential form as
\begin{equation}
\begin{split}
	\mathcal{L}(s)=&\,\exp\Bigg\{-s\sigma_\mathrm{n}^2-\sum_{j}\lambda_{\mathrm{t},j}\times\\
	&\,\int_{\mathbb{R}^2}\left(1-\mathbb{E}_{g_{x,j}}[\exp(-sg_{x,j}\Vert x\Vert^{-\alpha})]\right)\mathrm{d}x\Bigg\}\\
	=&\,\exp\{\eta(s)\},\label{Ls3}
	\end{split}
\end{equation}
where $\eta(s)$ is the exponent of the Laplace transform $\mathcal{L}(s)$.
 First, the recursive relations between $n$-th derivatives of the Laplace transform are illustrated in the following lemma.
\begin{lemma}\label{lem1}
Define $x_n=\frac{(-s)^n}{n!}\mathcal{L}^{(n)}(s)$, we have
\begin{equation}
x_n=\sum_{i=0}^{n-1}\frac{n-i}{n}q_{n-i}x_i,\quad q_k=\frac{(-s)^k}{k!}\eta^{(k)}(s).\label{nd}
\end{equation}
\end{lemma}
\begin{IEEEproof}
	See Appendix \ref{AA}.
\end{IEEEproof}
The calculation of the $n$-th derivatives commonly appears in the performance analysis of multi-antenna systems. However, direct computation leads to messy expressions \cite{5351444}. In contrast, the recursive relations in Lemma \ref{lem1}  enable us to express the $n$-th derivatives of $\mathcal{L}(s)$ in a delicate way, which leads to a compact matrix form of the coverage probability, as given in the following theorem.
\begin{theorem}($\ell_1$-Toeplitz Matrix Representation of the Coverage Probability)\label{th1} The coverage probability \eqref{coverageprob} is given by
	\begin{equation}
	p_\mathrm{c}(\gamma)=
	\int_0^\infty f_{r_0}(r)\left\Vert\exp\left\{\mathbf{Q}_M(r)\right\}\right\Vert_1\mathrm{d}r,\label{frameexpr}
	\end{equation}
	where $\mathbf{Q}_M$ is an $M\times M$ lower triangular Toeplitz matrix
	\begin{equation}
	\mathbf{Q}_M=\left[{\begin{IEEEeqnarraybox*}[][c]{,c/c/c/c/c,}
		q_0&{}&{}&{}&{}\\
		q_1&q_0&{}&{}&{}\\
		q_2&q_1&q_0&{}&{}\\
		\vdots &{}&{}& \ddots &{}\\
		q_{M-1}&\cdots&q_2& q_1 &q_0
		\end{IEEEeqnarraybox*}} \right].\label{topmatrix}
	\end{equation}
The nonzero entries of $\mathbf{Q}_M$ are determined by \eqref{nd}.
\end{theorem}
\begin{IEEEproof}
	See Appendix \ref{AA}.
\end{IEEEproof}

\newcounter{TempEqCnt}                         
\setcounter{TempEqCnt}{\value{equation}} 
\setcounter{equation}{9-1}
\begin{table*}[htbp]\small
	\centering
	\caption{Key parameters for different network settings}
	\begin{tabular}{|l|c|c|c|c|}
		\hline
		& \multicolumn{1}{c|}{\textbf{Multi-antenna}} & \multicolumn{1}{c|}{\textbf{Signal channel}} & \multicolumn{1}{c|}{\textbf{Interference}} & \multirow{2}[0]{*}{\textbf{Point process of the}} \\
		& \multicolumn{1}{c|}{\textbf{transmission} } & \multicolumn{1}{c|}{\textbf{gain ($g_{x_0}$)}} & \multicolumn{1}{c|}{\textbf{channel gain ($g_x$)}} & \multirow{2}[0]{*}{\textbf{interfering transmitters} $\Phi^\prime$}   \\
		& \multicolumn{1}{c|}{\textbf{technique}} & \multicolumn{1}{c|}{\textbf{distribution}} & \multicolumn{1}{c|}{\textbf{distribution}} &  \\\hhline{|=|=|=|=|=|}
		\textbf{Single-Antenna} & \multirow{2}[0]{*}{\diagbox[dir=SW,height=2.2em,width=8.5em]{}{}} & \multirow{2}[0]{*}{$\mathrm{Gamma}(1,1)$} & \multirow{2}[0]{*}{$\mathrm{Exp}(1)$} & \multirow{2}[0]{*}{$\mathcal{P}(r_0,\infty)$ with density $\lambda_\mathrm{t}$} \\
		\textbf{Networks }\cite{6042301} &       &       &       &  \\\hline
		\textbf{Throughput and Energy} & \multirow{2}[0]{*}{MRT} & \multirow{2}[0]{*}{$\mathrm{Gamma}(N_\mathrm{t},1)$} & \multirow{2}[0]{*}{$\mathrm{Exp}(1)$} & \multirow{2}[0]{*}{$\mathcal{P}(r_0,\infty)$ with density $\lambda_\mathrm{t}$} \\
		\textbf{Efficiency Analysis }\cite{6775036} &       &       &       &  \\\hline
		\textbf{Interference} & \multirow{2}[0]{*}{ZF beamforming} & $\mathrm{Gamma}$ & $g_{x,1}\sim\mathrm{Exp}(1)$& $\Phi^\prime_1$: $\mathcal{P}(r_0,\mu r_0)$ with density $\varepsilon\lambda_\mathrm{t}$ \\
		\textbf{Coordination }\cite{7038201} &       &$(\max(N_\mathrm{t}-K_{x_0},1),1)$       & $g_{x,2}\sim\mathrm{Exp}(1)$      & $\Phi^\prime_2$: $\mathcal{P}(\mu r_0,\infty)$ with density $\lambda_\mathrm{t}$ \\\hline
		\textbf{$K$-tier Multiuser} & \multirow{2}[0]{*}{SDMA} & \multirow{2}[0]{*}{$\mathrm{Gamma}(M_k-U_k+1,1)$} & \multirow{2}[0]{*}{$g_{x,j}\sim\mathrm{Gamma}(U_j,1)$} & \multirow{2}[0]{*}{$\Phi^\prime_j$: $\mathcal{P}_j(r_j,\infty)$ with density $\lambda_{\mathrm{t},j}$} \\
		\textbf{MIMO HetNets }\cite{7412737} &       &       &       &   \\\hline
		\textbf{Physical Layer Security} & Jamming \& & \multirow{2}[0]{*}{$\mathrm{Gamma}(N_{x_0},1)$} & {$g_{x,1}\sim\mathrm{Gamma}(N_x,1)$} & \multirow{2}[0]{*}{See Section \ref{IVA}} \\
		\textbf{Aware Networks} &   ZF beamforming    &       &   $g_{x,2}\sim\mathrm{Exp(1)}$    &  \\\hline
		\textbf{Millimeter-wave} & {Analog} & \multirow{2}[0]{*}{$\mathrm{Gamma}(M,1/M)$} & \multirow{2}[0]{*}{\eqref{sinsin}} & \multirow{2}[0]{*}{See Section \ref{IVB}} \\
		\textbf{Networks} & beamforming      &       &       &  \\\hline
	\end{tabular}%
	\label{table1}%
	\begin{tablenotes}
		\item * $\mathcal{P}(a,b)$ denotes a PPP within a ring with inner diameter $a$ and outer diameter $b$.
	\end{tablenotes}
\end{table*}
\setcounter{TempEqCnt}{\value{equation}} 
\setcounter{equation}{11-1}
	\begin{figure*}
	\begin{equation}\label{exampleeq}
	q_{k,i}=\frac{1}{P_k^\delta B_k^\delta}\sum_{j=1}^K\lambda_jP_j^\delta B_j^\delta\frac{\Gamma(U_j+i)}{\Gamma(U_j)\Gamma(i+1)}\frac{\delta}{i-\delta}\left(\frac{U_kB_k}{U_jB_j}\gamma\right)^i\times {}_2F_1\left(i-\delta,U_j+i;i+1-\delta;-\frac{U_kB_k}{U_jB_j}\gamma\right)
	\end{equation} \hrulefill
\end{figure*}
\setcounter{equation}{\value{TempEqCnt}}
Compared to the complicated approximations in \cite{4712724,5673756,5351444,6932503}, the $\ell_1$-Toeplitz matrix representation in \eqref{frameexpr} provides a much more compact form for the coverage probability. More importantly, it enables us to leverage various powerful tools from linear algebra, especially some nice properties of the lower triangular Toeplitz matrix, to provide insightful design guidelines for further network optimization. Such properties in the setting of small cell networks can be found in \cite{6775036}.

\subsection{Single-Antenna vs. Multi-Antenna Networks}
The proposed framework incorporates the single-antenna network \cite{6042301} as a simple special case. Assuming Rayleigh fading, the signal channel gain is exponentially distributed in the single-antenna case, i.e., $M=\theta=1$. Then, the expression \eqref{frameexpr} in Theorem \ref{th1} can be simplified as
\begin{equation}
	p_\mathrm{c}(\gamma)=
\int_0^\infty f_{r_0}(r)\mathcal{L}(s)\mathrm{d}r,
\end{equation}
which is exactly the same as the classic result in \cite[Equation 2]{6042301}. Note that, for single-antenna networks, the main task to derive the coverage probability is to manipulate the Laplace transform $\mathcal{L}(s)$. It has been shown in \cite{6042301} that, under various assumptions for the interference channel gain $g$ and different point processes of concerned interfering transmitters $\Phi^\prime$, $\mathcal{L}(s)$ (equivalently $\eta(s)$) can be derived into closed forms. This also creates the possibility to express the coverage probability in a closed form or a simple integral expression.

When it comes to multi-antenna networks, Theorem \ref{th1} is compatible with any specific form of $\eta(s)$ as long as the Laplace transform can be expressed as $\mathcal{L}(s)=\exp\{\eta(s)\}$.
Furthermore, with the gamma distributed signal channel gain, the only additional task compared to single-antenna networks is to calculate $M-1$ derivatives of $\eta(s)$, which will not introduce much computational complexity and thus maintains the tractability. This means that many manipulation tricks and steps developed for single-antenna networks can be transplanted to the multi-antenna case. The tractability and effectiveness of the proposed framework will be firstly illustrated in Section \ref{IIC} with some existing results as special cases, and then will be further demonstrated in Section \ref{III} via developing new analytical results.

\subsection{Examples}\label{IIC}
When applying Theorem \ref{th1} to specific multi-antenna networks, the only parameters to be determined are the nonzero entries $\{q_i\}_{i=0}^{M-1}$ in the matrix $\mathbf{Q}_M$. Thus, there are two main steps when applying the proposed framework:
\begin{itemize}
	\item First, we derive the Laplace transform $\mathcal{L}(s)$ for the given distribution of $g$ and the specific point process for the interfering transmitters $\Phi^\prime$.
	\item Then, we calculate the $n$-th ($1\le n\le M-1$) derivatives of the exponent $\eta(s)$ of the Laplace transform to compose $\{q_i\}_{i=0}^{M-1}$ in the matrix $\mathbf{Q}_M$ according to \eqref{nd}.
\end{itemize}  
Following is an example, which provides a closed-form expression for $\{q_i\}_{i=0}^{M-1}$ (also for $p_c(\gamma)$) in a general multiuser MIMO HetNet.
\begin{example} For a general $K$-tier multiuser MIMO HetNet with SDMA, as considered in \cite{7412737}, the
coverage probability can be expressed in a closed form as
\begin{equation}
p_c(\gamma)=\sum_{k=1}^K\left\Vert\mathbf{Q}^{-1}_{M_k-U_k+1}\right\Vert_1.
\end{equation}
The corresponding $\{q_{k,i}\}_{i=1}^{M_k-U_k}$ are provided in \eqref{exampleeq}, where $\delta=\frac{2}{\alpha}$ and ${}_2 F_1 \left(a,b;c;z\right)$ is the Gauss hypergeometric function \cite{zwillinger2014table}. 
\end{example}
\noindent Thanks to the proposed framework, this result is in a much more compact form than existing ones, and thus is amenable for further system analysis and optimization. Moreover, it is applicable to general multiuser MIMO HetNets.

As mentioned before, Theorem \ref{th1} is a generalization of our previous results in \cite{6775036,7038201,7412737}. The corresponding distributions of the channel gains and the point processes of interfering transmitters are listed in Table \ref{table1}\footnote{The physical meanings of the notations can be found in corresponding papers \cite{6775036,7038201,7412737}, and the parameters for two examples in the next section are also provided.}. By calculating the $n$-th derivatives of $\eta(s)$ and substitute them into the Toeplitz matrix, Theorem \ref{th1} specializes to the analytical results therein. Such tractable expressions yield lots of system design insights, as specified below.
\begin{itemize}
	\item In \cite{6775036}, it was analytically shown that the network throughput scales with the BS density first linearly, then logarithmically, and finally converge to a constant. 
	The energy efficiency will first increase and then decrease when increasing BS density/antenna size.
	\item In \cite{7038201}, a tractable coverage probability expression was derived with the proposed user-centric intercell interference nulling strategy, based on which the optimal intercell interference range was determined to further improve the network performance. 
	\item Trade-off between ASE and link reliability in multiuser MIMO HetNets was studied in \cite{7412737}. Analytical results for ASE and coverage probability were given, which were incorporated in an efficient algorithm to find the optimal BS density that achieves the maximum ASE while guaranteeing a certain link reliability. 
\end{itemize}


\section{Applications of the Proposed Framework}\label{III}
In this section, we will apply the proposed analytical framework to two newly emerging paradigms of multi-antenna networks.
With more and more mobile devices connected to the network, information security becomes one primary concern in dense wireless networks. Meanwhile, the reduced coverage requirement of dense networks makes is possible to exploit abundant bandwidth at mmWave bands.
In this section, we will analytically investigate physical layer security aware networks and mmWave networks, in both of which multi-antenna transmissions play a critical role.

\subsection{Physical Layer Security Aware Networks}\label{IVA}
While jamming is an effective way to enhance the network secrecy performance \cite{6587514}, interference nulling is important to suppress co-channel interference in dense networks \cite{7038201}, both of which rely on multi-antenna transmissions.
In this part, we will analytically find the optimal balance between jamming and interference nulling.

\subsubsection{Network Model}
We consider an ad hoc network consisting of legitimate nodes and eavesdroppers. The legitimate transmitters are modeled as a homogeneous PPP $\Phi$ with density $\lambda_{\rm t}$. Each transmitter is equipped with $N_\mathrm{t}$ antennas and has an intended receiver at a fixed distance $r_0$ in a random direction.
The passive eavesdroppers also form a homogeneous PPP with density $\lambda_\mathrm{e}$, which is independent to $\Phi$. 

\subsubsection{Joint Jamming and Interference Nulling}
We propose a joint jamming and interference nulling scheme. To avoid strong interference and possible strong jamming signals from nearby transmitters, each legitimate receiver requests interference nulling from the interfering transmitters within a distance $d_0$, called the \emph{coordination range}. Denote the number of requests received by the transmitter located at $x$ by $K_x$, which is random due to the random node locations, and it is possible that $K_x \geq N_\mathrm{t}$. Due to the limited spatial degrees of freedom, each transmitter can handle at most $N_\mathrm{t}-1$ requests. If a transmitter receives $K_x \geq N_\mathrm{t}$ requests, we assume it will randomly choose $N_\mathrm{t}-1$ receivers to suppress interference. 

After each transmitter determines the interference nulling targets, the transmitter will perform jamming aided beamforming at the subspace which is orthogonal to its intended channel and the channels to the $\min\left(K_x,N_\mathrm{t}-1\right)$ receivers. Therefore, the jamming signal sent by the transmitter will not affect its own receiver and the other $\min\left(K_x,N_\mathrm{t}-1\right)$ receivers. But it will degrade the quality of service of the eavesdroppers and all the other receivers.
We denote $N_x = N_\mathrm{t} - \min\left(K_x,N_\mathrm{t}-1\right)$ as the total number of transmitted streams. Generally, increasing the coordination range $d_0$ will suppress more nearby interference but less jamming signals will be transmitted, which leads to a trade-off between the interference nulling and jamming.

\subsubsection{Connection Outage Probability}
Consider the typical receiver at the origin, whose transmitter locates at $x_0$ and receives $K_{x_0}$ requests of interference nulling. Then, based on the proposed scheme, the SIR can be given similar to \eqref{coverageprob}, and the needed parameters in the framework are listed as follows.
\begin{itemize}
	\item Signal channel gain: $g_{x_0}\sim\mathrm{Gamma}(N_{x_0},1)$;
	\item Point processes of the interfering transmitters and corresponding interference channel gains:\\
	$\Phi^\prime_\mathrm{out}=\mathcal{P}(d_0,\infty)$ with $g_x\sim\mathrm{Gamma}(N_x,1)$;\\
	$\Phi^\prime_\mathrm{in}=\mathcal{P}(0,d_0)$ with $g_x\sim\mathrm{Exp}(1)$.
\end{itemize}
The connection outage probability $p_\mathrm{co}$ \cite{6587514}, defined as the probability that the SIR of a typical receiver is
below a certain threshold $\gamma_l$, is presented in the following proposition.
\setcounter{equation}{12-1}
\begin{prop} \label{Thm:pco}
	The connection outage probability of the typical legitimate receiver is given by
	\begin{equation} \label{eq:pco_Num}
	p_{{\rm co}}=
	1-\sum_{N_{x_{0}}=1}^{N_\mathrm{t}}p_{N}\left(N_{x_{0}}\right)p_{{\rm co}}\left(N_{x_{0}}\right),
	\end{equation}
	where 
	\begin{equation}
	p_{N}\left(n\right)=
	\begin{cases}
	\frac{\left(\pi d_{0}^{2}\lambda_\mathrm{t}\right)^{N_\mathrm{t}-n}}{\left(N_\mathrm{t}-n\right)!}e^{-\pi d_{0}^{2}\lambda_\mathrm{t}}, & n=2,3,\cdots,N_\mathrm{t},\\
	1-\sum_{i=2}^{N_\mathrm{t}}p_{N}\left(i\right), & n=1,
	\end{cases}
	\end{equation}
	\begin{equation} \label{eq:pcoN_rgeq0}
	p_{{\rm co}}\left(N_{x_{0}}\right)=1-
	\left\Vert \exp\left\{-\pi\lambda_\mathrm{t}d_{0}^{2}\left[\mathbf{Q}_{N_{x_{0}}}-\mathbf{I}_{N_{x_0}}\right]\right\}\right\Vert _{1}.
	\end{equation}
	The nonzero elements of $\mathbf{Q}_{N_{x_0}}$ are given by \eqref{eq:qi_r0}.
	\begin{figure*}
		\begin{equation} \label{eq:qi_r0}
		q_k=\sum_{n=1}^{N_\mathrm{t}}p_{N}\left(n\right) \frac{\Gamma\left(n+k\right)}{\Gamma\left(n\right)\Gamma\left(k+1\right)} \frac{\delta}{\delta-k}  \left[\left(\frac{r_0}{d_0}\right)^{\alpha}\gamma_l  \frac{N_{x_{0}}}{n}\right]^{k}\times{}_{2}F_{1}\left(k-\delta,k+n;k+1-\delta; -\left(\frac{r_0}{d_0}\right)^{\alpha}\gamma_l \frac{N_{x_{0}}}{n} \right)
		\end{equation}
		\setcounter{TempEqCnt}{\value{equation}} 
		\setcounter{equation}{23-1}
			\begin{equation}\label{eq29}
		\hat{q}_k=\frac{2\lambda\Gamma\left(k+\frac{1}{2}\right)\Gamma(M+k)\gamma^k}{\sqrt{\pi}d(k!)^2(\alpha k-2)\Gamma(M)}
		\Bigg[y_k\left(-\gamma\right)-(\pi\lambda_\mathrm{t})^2R^{2-\alpha k}\int_0^{R^2}e^{-\pi\lambda_\mathrm{t}r}r^\frac{\alpha k}{2}J_k\left(-\frac{\gamma}{R^{\alpha}}r^\frac{1}{\delta}\right)\mathrm{d}r\Bigg]\\
		\end{equation}
		\hrulefill
	\end{figure*}
\setcounter{equation}{\value{TempEqCnt}}
\end{prop}
\begin{IEEEproof}
	The proof is omitted due to space limitation.
\end{IEEEproof}	

\subsubsection{Secrecy Outage Probability}
Consider the eavesdropper located at $z$, the received SIR of this eavesdropper is given by
\begin{equation} \label{eq:SIR_e_def}
{\rm SIR}_{e,z}=\frac{\frac{P_{t}}{N_{x_{0}}}\tilde{g}_{0}\left\Vert x_{0}-z\right\Vert ^{-\alpha}}{\frac{P_{t}}{N_{x_{0}}}\tilde{g}_{x_{0}}\left\Vert x_{0}-z\right\Vert ^{-\alpha}+\sum_{x\in\Phi\backslash\left\{ x_{0}\right\} }\frac{P_{t}}{N_{x}}\tilde{g}_{x}\left\Vert x-z\right\Vert ^{-\alpha}},
\end{equation}
where the corresponding parameters in the framework are given as follows.
\begin{itemize}
\item Signal channel gain: $\tilde{g}_{0}\sim{\rm Gamma}\left(1,1\right)$;
\item Point processes of the interfering transmitters and corresponding interference channel gains:\\
Point $x_0$ with $\tilde{g}_{x_{0}}\sim{\rm Gamma}\left(N_{x_{0}}-1,1\right)$;\\
$\Phi^\prime=\mathcal{P}(0,\infty)\backslash\{x_0\}$ with $\tilde{g}_{x}\sim{\rm Gamma}\left(N_x,1\right)$.
\end{itemize}

Secrecy outage probability is defined as the probability that the SIR of at least one eavesdropper is above a certain threshold $\gamma_e$ \cite{6587514}.
A tight upper bound of the secrecy outage probability $p_{\rm so}$ is presented in Proposition \ref{Thm:pso}, as the exact expression of $p_{\rm so}$ is intractable. The proof is committed due to space limitation. 

\begin{prop} \label{Thm:pso}
	The secrecy outage probability $p_{\rm so}$ is upper bounded as
	\begin{equation} \label{eq:pso_UB}
	p_{\rm so} \leq   1- \sum_{N_{x_{0}}=1}^{N_\mathrm{t}} p_{N}\left(N_{x_{0}}\right)e^{ -\frac{\lambda_\mathrm{e}}{\lambda_\mathrm{t}} \frac{\left(1+{\gamma}_{e}\right)^{1-N_{x_{0}}} {\gamma}_{e}^{-\delta}N_{x_0}^{-\delta}} {\Gamma\left(1-\delta\right)\sum_{n=1}^{N_\mathrm{t}}p_{N}\left(n\right) \frac{\Gamma\left(n+\delta\right)}{\Gamma\left(n\right)n^{\delta}}} }.
	\end{equation}
\end{prop}

The secrecy transmission capacity is adopted as the main performance metric, which is defined as the achievable rate of
confidential messages per unit area with given connection and secrecy outage constraints.
To obtain the secrecy transmission capacity $C_s$ for a fixed $d_0$, we firstly find the SIR threshold $\gamma_l^{\rm th}$ satisfying the equation $p_{\rm co}=\mu$ using \eqref{eq:pco_Num}, and then find the SIR threshold $\gamma_e^\mathrm{th}$ satisfying the equation $p_{\rm so}=\epsilon$ using \eqref{eq:pso_UB}. Thus, the secrecy transmission capacity can be written as
\begin{equation}
C_s \left(d_0\right) = \left(1-\mu\right)\lambda_\mathrm{t}\left[\log_{2}\left(\frac{1+\gamma_{l}^{{\rm th}}}{1+\gamma_{e}^{{\rm th}}}\right)\right]^{+}.
\end{equation}
The design goal is to find the optimal $d_0$ to maximize the secrecy transmission capacity.
\begin{figure}
	\centering\includegraphics[height=5.5cm]{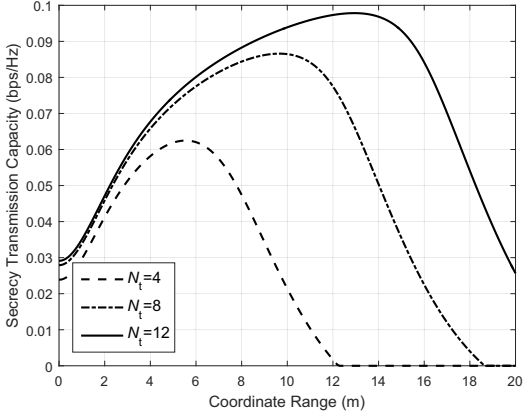}
	\caption{The secrecy transmission capacity with different $d_0$, with $\lambda_\mathrm{t}=10^{-2}$ m$^{-2}$, $\lambda_\mathrm{e}=10^{-3}$ m$^{-2}$, $r_0=1$ m, and $\alpha=4$. The connection outage constraint is $\mu=0.1$ while the secrecy outage constraint is $\epsilon=0.01$. }\label{fig1}
\end{figure}

In Fig. \ref{fig1}, we show $C_s \left(d_0\right)$ as a function of $d_0$ according to the analytical results we derive. We find that there is an optimal $d_0$ for each curve.  The reason that increasing $d_0$ from $0$ can increase $C_s$ is that nearby interference is critical for the legitimate receivers. Thus, setting a protecting zone for each receiver could significantly improve the performance of legitimate receivers. However, if $d_0$ is too large, $C_s$ will decrease, as a too large $d_0$ means each receiver requests many transmitters for interference nulling. While the performance improvement of the legitimate receiver is diminishing, the degrees of freedom left for each transmitter to send jamming noise will be small. Thus, the eavesdroppers will experience a better received SIR. Based on the tractable expressions of outage probabilities, we can easily find the optimal $d_0$. As $d_0=0$ corresponds to the special case without interference nulling, Fig. \ref{fig1} also shows that with the optimal $d_0$, the proposed scheme achieves significant performance gains over the scheme only based on jamming \cite{6587514}, which implies the importance of interference management in jamming assisted networks.

\subsection{Millimeter Wave Cellular Networks}\label{IVB}
In mmWave networks, directional antenna arrays are used both to combat huge path loss and to synthesize narrow beams, which also differentiates mmWave networks from conventional ones. However, how the directional arrays effect the network performance has not been fully understood. In this subsection, we adopt the proposed framework to investigate the role of directional antenna arrays in mmWave cellular networks.
\subsubsection{Network Model}
We assume that mobile users are distributed as a homogeneous PPP, which is independent of $\Phi$, and each user is associated with the nearest BS. In mmWave cellular networks, one unique characteristic is the blockage effect.
It has been pointed out in \cite{6932503} that non-line-of-sight (NLOS) signals and NLOS interference are negligible in dense mmWave networks. Hence, we will focus on the analysis where the typical user is associated with a LOS BS and the interference stems from LOS BSs, which are distributed according to the PPP $\mathcal{P}(r_0,R)$ if we adopt the \textit{line-of-sight (LOS) Ball} blockage model \cite{6932503}, where $R$ is the LOS radius.

\subsubsection{Impact of Directional Arrays}
With a uniformly random single path (UR-SP) channel model and analog beamforming, the SINR at the typical user can be expressed in the same form as \eqref{coverageprob}, with the required parameters listed as follows \cite{7564903}.
\begin{itemize}
	\item Signal channel gain: $g_{x_0}=\left|\rho_{0}\right|^2\sim\mathrm{Gamma}\left(M,\frac{1}{M}\right)$;
	\item Point process of the interfering transmitters $\Phi^\prime=\mathcal{P}(r_0,R)$ and corresponding interference channel gains:
\begin{equation}\label{sinsin}
g_x=\left|\rho_x\right|^2\frac{\sin^2\left(\frac{d}{\lambda}\pi \varphi_x\right)}{N_\mathrm{t}^2\sin^2\left(\frac{d}{\lambda}\pi \varphi_x\right)}\triangleq \left|\rho_x\right|^2 G_\mathrm{act}(\varphi_x),
\end{equation}
\end{itemize}where $\varphi_x$ are independent uniformly distributed random variables over $\left[-1,1\right]$. The array gain function in \eqref{sinsin} is referred as the \emph{actual antenna pattern}.
The main difficulty in analyzing the distribution of SINR is the complicated distribution of $g_x$. In oder to obtain a tractable analysis result, we adopt an approximation for the array gain function $G_\mathrm{act}$, which is referred as the \textit{cosine antenna pattern}
\begin{equation}\label{approxpattern}
G_\mathrm{cos}(x)=
\begin{cases}
\cos^2\left(\frac{\pi N_\mathrm{t}}{2}x\right)&|x|\le\frac{1}{N_\mathrm{t}},\\
0&\text{otherwise}.
\end{cases}
\end{equation}
With the approximated cosine antenna pattern and Theorem 1, a lower bound for the coverage probability can be provided to present the impact of directional antenna arrays, as shown in the following proposition.

\begin{prop}\label{coro2}
	The coverage probability is tightly lower bounded by
	\begin{equation}\label{b33}
	p_\mathrm{c}^{\cos}(t)\ge \left(1-e^{-\pi\lambda_\mathrm{t}R^2}\right)e^{\beta_0t}\left(1+\sum_{n=1}^{M-1}\beta_nt^n\right),
	\end{equation}
	which is a non-decreasing concave function of the array size $N_\mathrm{t}$, and $t=\frac{1}{N_\mathrm{t}}$,
	\begin{equation}
	\beta_n=
	\begin{dcases}
	\frac{\hat{q}_0}{1-e^{-\pi\lambda_\mathrm{t}R^2}}&n=0,\\
	\frac{\left\Vert \left(\hat{\mathbf{Q}}_M-\hat{q}_0\mathbf{I}_M\right)^n\right\Vert_1}{n!\left(1-e^{-\pi\lambda_\mathrm{t}R^2}\right)}&n\ge1.
	\end{dcases}
	\end{equation}
	The nonzero entries in $\hat{\mathbf{Q}}_M$ are determined by \eqref{eq29}, and
\setcounter{equation}{23}
	\begin{equation}
	\begin{split}
	y_k(x)=&J_k(x)\left[1-e^{-\pi\lambda_\mathrm{t}R^2}\left(1+\pi\lambda_\mathrm{t}R^2\right)\right]\\
	&+\mathbf{1}(k=0)\left(
	\pi\lambda_\mathrm{t}R^2-1+e^{-\pi\lambda_\mathrm{t}R^2}\right).
	\end{split}
	\end{equation}
		where
	\begin{equation}
	J_k\left(x\right)={}_3F_2\left(k+\frac{1}{2},k-\delta,k+M;k+1,k+1-\delta;x\right),
	\end{equation}
	with ${}_3F_2(a_1,a_2,a_3;b_1,b_2;z)$ denoting the generalized hypergeometric function \cite{zwillinger2014table} and $\mathbf{1}(\cdot)$ being the indicator function.
\end{prop}
\begin{IEEEproof}
	The proof is omitted due to space limitation.
\end{IEEEproof}	

From Proposition \ref{coro2}, we discover that increasing the directional antenna array size will definitely benefit the coverage probability, and the concavity means that the benefits on the coverage brought from leveraging more antennas will gradually diminish with the increasing antenna size. Moreover, we see that the lower bound is a product of an exponential function and an $M$-degree polynomial function with respect to the inverse of the array size $t$. For the special case that $M=1$, i.e., the Rayleigh fading channel, the lower bound will reduce to an exponential one.
\begin{figure}[tbp]
	\centering\includegraphics[height=5.5cm]{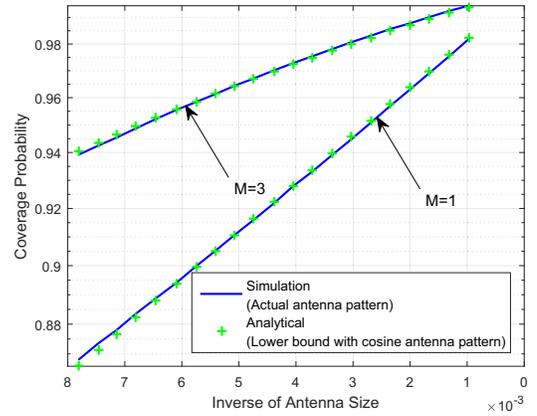}
	\caption{Impact of antenna arrays in mmWave cellular networks when $R=200$ m, $\gamma=5$ dB, $P_\mathrm{t}=1$ W, $\lambda_\mathrm{t}=10^{-3}$ m$^{-2}$, $\beta=-61.4$ dB, and $\alpha=2.1$.}\label{fig2}
\end{figure}

Fig. \ref{fig2} demonstrates that the analytical result in Proposition \ref{coro2} well matches the simulation result. The performance gain with a larger array size is mainly because increasing the array size will narrow the interference beam, which reduces the probability that the interferers direct the main lobes towards the typical user.
In addition, as stated before, when $M=1$, the lower bound \eqref{b33} will reduce to an exponential one, which is linear in the logarithm scale as shown in Fig. \ref{fig2}. When the Nakagami parameter $M$ increases, the polynomial term will take effect to make the lower bound to be a concave one.

\section{Conclusions}
This paper proposed a unified analytical framework based on the $\ell_1$-Toeplitz matrix representation for coverage analysis of dense multi-antenna networks. A tractable expression for a general network model was firstly derived. Two examples, i.e., the security aware wireless networks and mmWave networks, were then provided to demonstrate the generality and effectiveness of the proposed framework. Overall, this paper provided a powerful toolbox for the evaluation and design of various dense multi-antenna wireless networks, which will find ample applications.

\appendices
\section{Proof of Lemma \ref{lem1} and Theorem \ref{th1}}\label{AA}
Defining $x_n=\frac{(-s)^n}{n!}\mathcal{L}^{(n)}(s)$, the coverage probability \eqref{eq4} can be expressed as
\begin{equation}\label{25}
p_\mathrm{c}(\gamma)=\mathbb{E}_{r_0}\left[\sum_{n=0}^{M-1}x_n\right],
\end{equation}
where $x_0=\mathcal{L}(s)=\exp\{\eta(s)\}$ is given in Lemma \ref{lem1}. Next, we will express $x_n$ in a recursive form. It is obvious that $\mathcal{L}^{(1)}(s)=\eta^{(1)}(s)\mathcal{L}(s)$, and according to the formula of Leibniz for the $n$-th derivative of the product of two functions, we have
\begin{equation}
\mathcal{L}^{(n)}(s)=\frac{\mathrm{d}^{n-1}}{\mathrm{d}s}\mathcal{L}^{(1)}(s)=\sum_{i=0}^{n-1}{{n-1}\choose i} \eta^{(n-i)}(s)\mathcal{L}^{(i)}(s),
\end{equation}
followed by
\begin{equation}
\frac{(-s)^n}{n!}\mathcal{L}^{(n)}(s)=\sum_{i=0}^{n-1}\frac{n-i}{n}\frac{(-s)^{(n-i)}}{(n-i)!}\eta^{(n-i)}(s)\frac{(-s)^i}{i!}\mathcal{L}^{(i)}(s).
\end{equation}
Therefore, the recursive relationship of $x_n$ is
\begin{equation}\label{recur}
x_n=\sum_{i=0}^{n-1}\frac{n-i}{n}q_{n-i}x_i,
\end{equation}
where
\begin{equation}
q_k=\frac{(-s)^k}{k!}\eta^{(k)}(s).
\end{equation}
This completes the proof of Lemma \ref{lem1}.

Then, we define two power series as follows to solve for $x_n$,
\begin{equation}\label{eq40}
Q(z)\triangleq\sum_{n=0}^\infty q_nz^n,\quad 
X(z)\triangleq\sum_{n=0}^\infty x_nz^n.
\end{equation}
Using the properties that $Q^{(1)}(z)=\sum_{n=0}^{\infty}nq_nz^{n-1}$ and $Q(z)X(z)=\sum_{n=0}^\infty\sum_{i=0}^nq_{n-i}x_iz^n$, from \eqref{recur}, we obtain the differential equation
\begin{equation}
X^{(1)}(z)=Q^{(1)}(z)X(z),
\end{equation}
whose solution is
\begin{equation}
X(z)=\exp\left\{Q(z)\right\}.\label{40}
\end{equation}
Therefore, according to \eqref{25}, \eqref{eq40}, and \eqref{40}, the coverage probability is given by
\begin{equation}
\begin{split}
p_\mathrm{c}(\gamma)&=\mathbb{E}_{r_0}\left[\sum_{n=0}^{M-1}x_n\right]=\mathbb{E}_{r_0}\left[\sum_{n=0}^{M-1}\frac{1}{n!}\left.{X^{(n)}(z)}\right|_{z=0}\right]\\
&=\mathbb{E}_{r_0}\left[\sum_{n=0}^{M-1}\frac{1}{n!}\frac{\mathrm{d}^n}{\mathrm{d}z^n}\left.{e^{Q(z)}}\right|_{z=0}\right].
\end{split}
\end{equation}
From \cite[Page 14]{henrici1974applied}, the first $M$ coefficients of the power series $e^{Q(z)}$ form the first column of the matrix exponential $\exp\{\mathbf{Q}_M\}$, whose exponent is given in \eqref{topmatrix}.

\bibliographystyle{IEEEtran}
\bibliography{bare_conf}
\end{document}